# Projections of Economic Impacts of Climate Change on Marine Protected Areas: Palau, the Great Barrier Reef, and the Bering Sea

Talya ten Brink[1]

[1] Department of Marine Affairs, University of Rhode Island, Kingston, RI 02881, USA

## Abstract

Climate change substantially impacts ecological systems. Marine species are shifting their distribution because of climate change towards colder waters, potentially compromising the benefits of currently established Marine Protected Areas (MPAs). Therefore, we demonstrate how three case study regions will be impacted by warming ocean waters to prepare stakeholders to understand how the fisheries around the MPAs is predicted to change. We chose the case studies to focus on large scale MPAs in i) a cold, polar region, ii) a tropical region near the equator, and iii) a tropical region farther from the equator. We quantify the biological impacts of shifts in species distribution due to climate change for fishing communities that depend on the Palau National Marine Sanctuary, the Great Barrier Reef Marine National Park Zone, and the North Bering Sea Research Area MPAs. We find that fisheries sectors will be impacted differently in different regions and show that all three regions can be supported by this methodology for decision making that joins sector income and species diversity.

## 1. Introduction

Marine protected areas (MPAs) are designed to protect the biodiversity of marine species and the local communities that depend upon them. Management measures of Marine Protected Areas are areas in the ocean that are protected for species. They provide many benefits, including tourism, recreation, watershed protection, biodiversity, education, research, consumptive benefits such protecting stocks for fishing and more (Angulo-Valdes and Hatcher, 2010). However, little is known about their effectiveness for industrial, artisanal, recreational, and subsistence fisheries as the temperatures of the ocean rise. As climate change reshapes marine habitats, species often decline or shift towards higher latitudes and deeper waters. These shifts compromise the envisioned benefits and management objectives of MPAs as species move outside the regulatory boundaries that were designed to protect them. A recent paper (Palacios-Abrantes et al., 2023) looks at the global impact of climate change on MPAs worldwide. However, this method has not been used on a smaller regional scale to predict the changes on single MPAs. To explore the impacts of climate change near large-scale MPAs we propose to analyze three case studies: Palau National Marine Sanctuary, Great Barrier Reef Marine National Park Zone, and the Northern Bering Sea Research Area. These three case studies offer different perspectives due to their climate and surrounding fisheries economic structures. For each case study we estimate the impacts of climate change on the MPA through two main parameters (biomass and fishing revenue), for four fisheries sectors: industrial, artisanal, subsistence, and recreational fisheries. These sectors are defined by the dataset, the Sea Around Us (Zeller et al., 2016). The industrial sector contains all gears towed intensively through the water column using engine power. The artisanal sector is assumed to operate only in domestic waters at a maximum of 50 km from the coast or to 200 m in depth, whichever comes first, and consists of small scale (i.e. gillnets and hand lines) and fixes gears, such as traps, resulting in catch sold commercially. It is also only for territories that have a permanent human population. The subsistence sector contains fisheries that are non-commercial fish that is consumed by the fisher or their family. The recreational sector contains fisheries that are caught for recreation or pleasure, and non-commercial because the fish caught are not sold.

The research question is:

1. How does climate change affect species biomass, catch, and revenue in Palau, the Great Barrier Reef, and the Northern Bering Sea?

## 2. Methods

### 2.1. Case selection

We investigate how climate change affects artisanal, subsistence, commercial, and recreational fisheries outcomes inside and adjacent to MPAs and what policy choices could best promote ecosystem and user outcomes. We have selected three large-scale MPAs: Palau National Marine Sanctuary (Palau), the Great Barrier Reef (GBR) Marine National Park Zone (Australia), and the North Bering Sea Research Area (USA). The motivation for the selection of these three large-scale MPAs is the variance in economic status, systems and climate, with each of the cases having artisanal and commercial fisheries, and recreational fishing stakeholders who rely on the long-term socio-economic benefits from the marine ecosystem MPA (Bering Sea: Huffines, 2018; GBR: Graham et al. 2003; McCook et al. 2010; Palau: PICRC, 2019; World Wildlife Fund, 2019). These case studies can serve as examples for other coastal nations that are considering implementing, maintaining, or modifying their regulations regarding use of marine resources.

Palau is a small-island country in the Pacific composed of approximately 300 islands. The islands form a presidential republic in free association with the United States since 1982 and have a total population of 17,862 (Gillet, 2017), an EEZ of approximately 600,000 km 2 , and approximately 500 km2 of land. The government established the Palau National Marine Sanctuary (PNMS) in 2015 and have been phasing in regulations in the sanctuary. The Palau National Marine Sanctuary Act (2015) is the sixth largest MPA in the world and in 2015, currently 80% of the Palau's EEZ operates as a no-take reserve and 20% as a domestic fishing zone (Palau National Marine Sanctuary Act (RPPL No. 9-49 of 2015).

The Great Barrier Reef Marine Park extends approximately 2,400 km along the northeastern coast of Australia in the state of Queensland. The no-take area of the Greater Great Barrier Reef, the Marine Park, was established in 1975 and was given UNESCO World Heritage status in 1981. The area encompasses approximately 552,000 km$^2$ and provides habitat for one of the most biodiverse marine species in the world, thereby offering opportunities for fisheries and tourism (Wynveen and Kyle 201). The Aboriginal and Torres Strait Islander people also have sustainably used the resources of the Great Barrier Reef for centuries. The Great Barrier Reef supports a variety of stakeholders including commercial fisheries, recreation, and indigenous populations, which target a range of species, including fish, sharks, crabs and prawns (Great Barrier Reef Marine Park Authority, 2019). In 2012, $193 million was generated through commercial fishing and aquaculture in the region (Great Barrier Reef Park Authority, 2013). Furthermore, recreational fishing has a much larger economic impact than the commercial sector (Prayaga et al, 2010), and puts additional fishing pressure on the Reef (Mapstone et al. 2004). However, 90% of recreational fishers comply with the regulations of the no-take zones (Arias & Sutton, 2013).

The North Bering Sea Research Area extends from the area of Bethel to the Cape Prince of Wales in Alaska, USA. The North Bering Sea National Marine Sanctuary encompasses 222,179 km$^2$. The North Bering Sea Research Area was established in 2008 as a scientific research MPA (Witherell & Woodby, 2005). The area provides habitat for diverse bird, fish and marine mammal species. It also plays an important role for distinct indigenous communities (Oceana, 2014). Commercial fishing pressure, while not currently intense, is expected to rise in the future with climate change as fish seek colder climates in a warming ocean (Whitehouse et al, 2021). Currently, the North Bering Sea Research Area is closed to bottom trawling, to test the hypothesis that intensive trawl fishing may create a local depletion of Pacific cod. So far, studies have found no differences in Pacific cod biomass between the intensively trawled areas and the un-trawled control areas (Witherell & Woodby, 2005). However, with commercial fish stocks in the southern Bering Sea moving northward, it is unclear whether the regulations would change in the future. The only species currently commercially harvested in the North Bering Sea is yellowfin sole (NOAA Fisheries, 2012). In many communities, subsistence fishing and hunting is of equal or greater importance than commercial fisheries.

## 2.2. Measuring economic outcomes of MPAs under climate change

### 2.2.1. Fisheries Data

Fisheries catch statistics compiled nationally and internationally systemically underreport certain fishing activities. Fisheries catch statistics often are biased towards including large scale industrial fisheries landings, instead of accounting for small-scale fisheries, including artisanal, subsistence and recreational sectors, and for ignoring discarded catch in overall counts. To solve this data challenge, the Sea Around Us has reconstructed fisheries catches (landings and discards) for all maritime countries in the world from 1950 to present (Pauly & Zeller, 2016; Zeller et al., 2016). In addition, they are spatially assigned to various scales used in fisheries management (regional fisheries management organization areas, exclusive economic zones (EEZs) and territorial waters, etc.) and allocated by species and habitat distributions to a global 0.5 by 0.5 scale. As mentioned earlier, the spatial allocation of data only allows small-scale fisheries (artisanal, subsistence, and recreational sectors) to operate within 50 nm of the coast, whereas industrial fleets are not restricted to this inshore fishing area (Zeller et al, 2016).

In addition to the catch data, partnerships have made possible extensive fisheries economics data that cover all fisheries globally. These data include the global ex-vessel price database (Tai et al., 2017), the cost of fishing database (Lam et al., 2011), fisheries subsidies data (Sumaila et al., 2010; Sumaila, et al., 2016), and country level expenditure on MPAs. The combination of these allow large-scale analyses of MPAs when combined with corollary data on the effects of MPAs on fish biomass and catches.

### 2.2.2 Climate Change Projections

Projected changes in the distribution of commercially important marine species under climate change were modeled for each of the MPA regions using a dynamic bioclimate envelope model (DBEM) (Cheung et al. 2009). Climate change variables were modeled from three earth system models: The Geophysical Fluid Dynamics Laboratory ESM 2M (GFDL) (www.gfdl.noaa.gov), the Institute Pierre Simon Laplace Climate Model 5 (IPSL-CM5) (www.http://icmc.ipsl.fr/), and the Max Planck Institute for Meteorology Earth System Model (MPI) (https://www.mpimet.mpg.de/en/science/models/). The model ran one IPCC-Representative Concentration Pathways: 8.5 (RCP8.5) representing a high emission scenario since it is the current 'business as usual' scenario (Nature Paper: Geographical limits to species-range shifts are suggested by climate velocity (Burrows et al., 2014; Moss et al. 2010).

The DBEM algorithm integrates ecophysiology, habitat suitability with spatial population dynamics of exploited fish and invertebrates to project shifts in biomass and maximum catch potential under climate change. The algorithm uses depth range, latitudinal range, habitat preferences and an index of species association with major habitat types to estimate changes in biomass distribution over a 0.5° latitude x 0.5° longitude grid of the world oceans (Cheung et al. 2009). For each grid cell and time step, the model simulates the species carrying capacity from sea surface temperature, salinity, oxygen content, sea ice extent (for polar species) and bathymetry. It then incorporates the intrinsic population growth, settled larvae, and net migration of adults from surrounding cells using an advection-diffusion-reaction equation. Finally, the model also simulates how changes in temperature and oxygen content would affect the growth of the individuals (Lam et al. 2016; Pauly and Cheung 2018).

We analyzed the grid cells immediately adjacent to a protected cell, i.e. grid cells in an MPA and directly adjacent to the MPA (Palacios-Abrantes et al, 2023) for both biomass and revenue. We calculated the biomass percentage change between two time periods, 2020 A.D. and midcentury (2020-2060) under RCP 8.5. Our data analysis was conducted with the statistical software R version 3.6.2 (2019-12-12).

## 3. Results

### 3.1. Spatial Representation of biomass change by midcentury (2020-2060 A.D.) for case studies

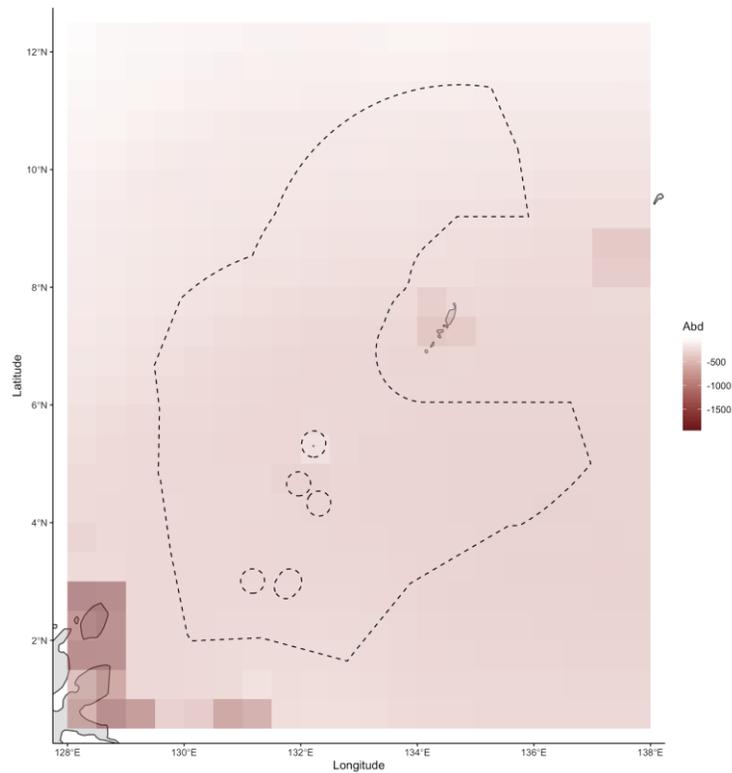

Figure 2a. Change in biomass aggregated across all species for the Palau MPA and surrounding area with a resolution of 0.5 by 0.5 degrees. RCP: 8.5 Dashed lines indicated MPA boundaries.

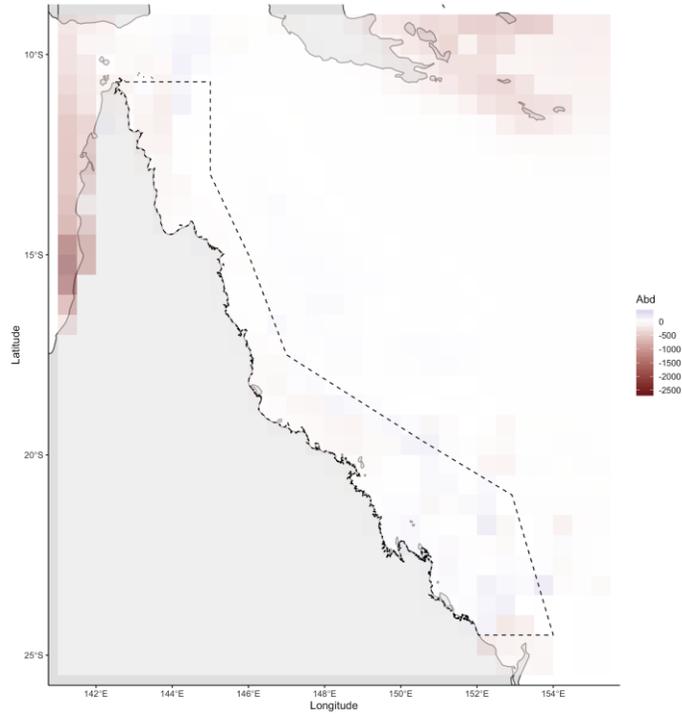

Figure 2b. Change in biomass aggregated across all species for the Great Barrier Reef MPA and surrounding area with a resolution of 0.5 by 0.5 degrees. RCP: 8.5 Dashed lines indicated MPA boundaries.

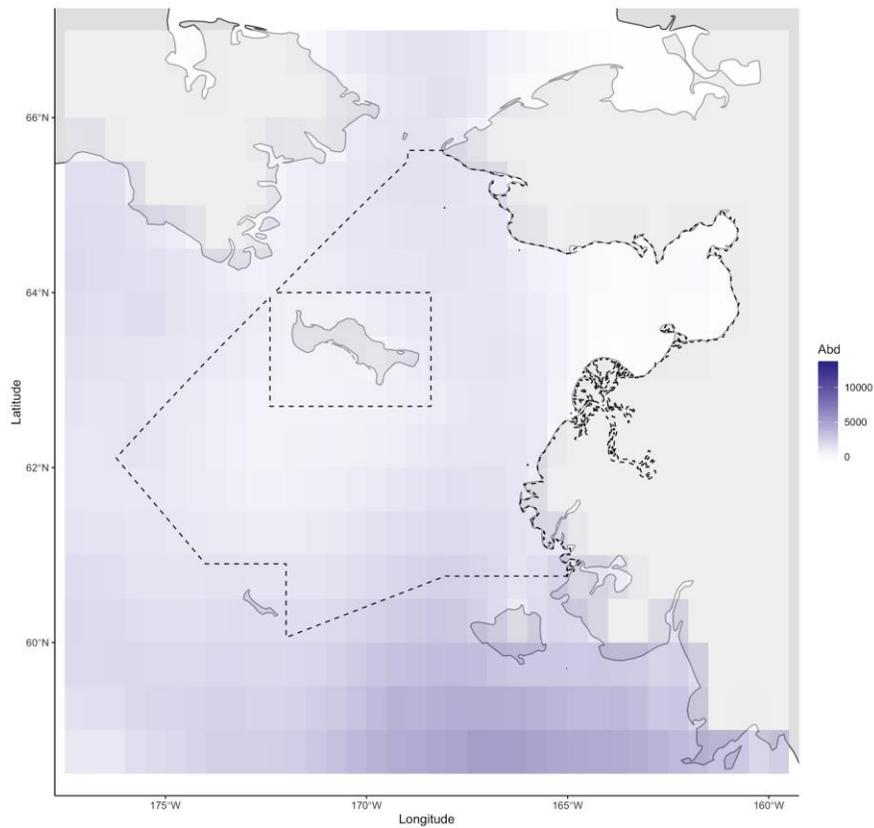

Figure 2c. Change in biomass aggregated across all species for the Bering Sea MPA and surrounding area with a resolution of 0.5 by 0.5 degrees. RCP: 8.5 Dashed lines indicated MPA boundaries.

## 3.2. Biomass for species change in all three case studies

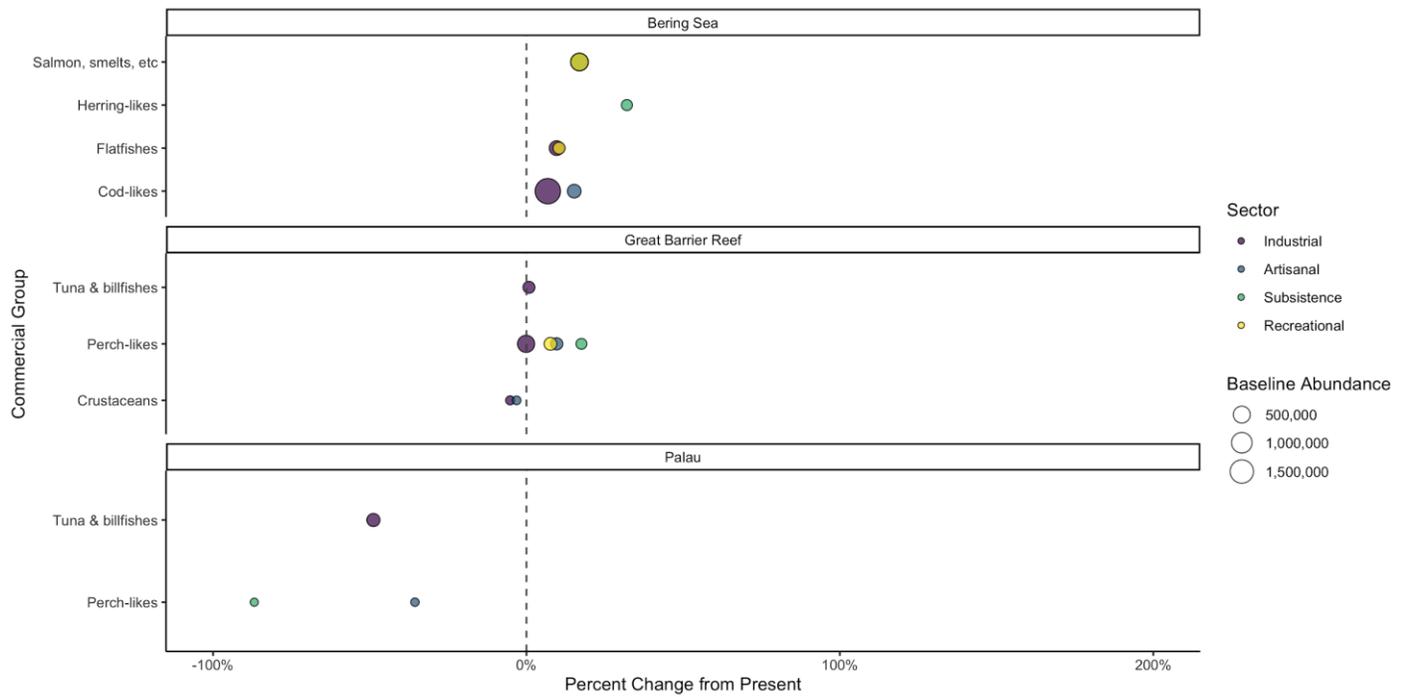

Figure 3. Percent change in biomass from present (1995-2014) to Mid-Century (2040-2060) for most important functional groups for each fishing sector. Bubble size indicates present biomass (tonnes).

## 3.2. Revenue for species change in all three case studies

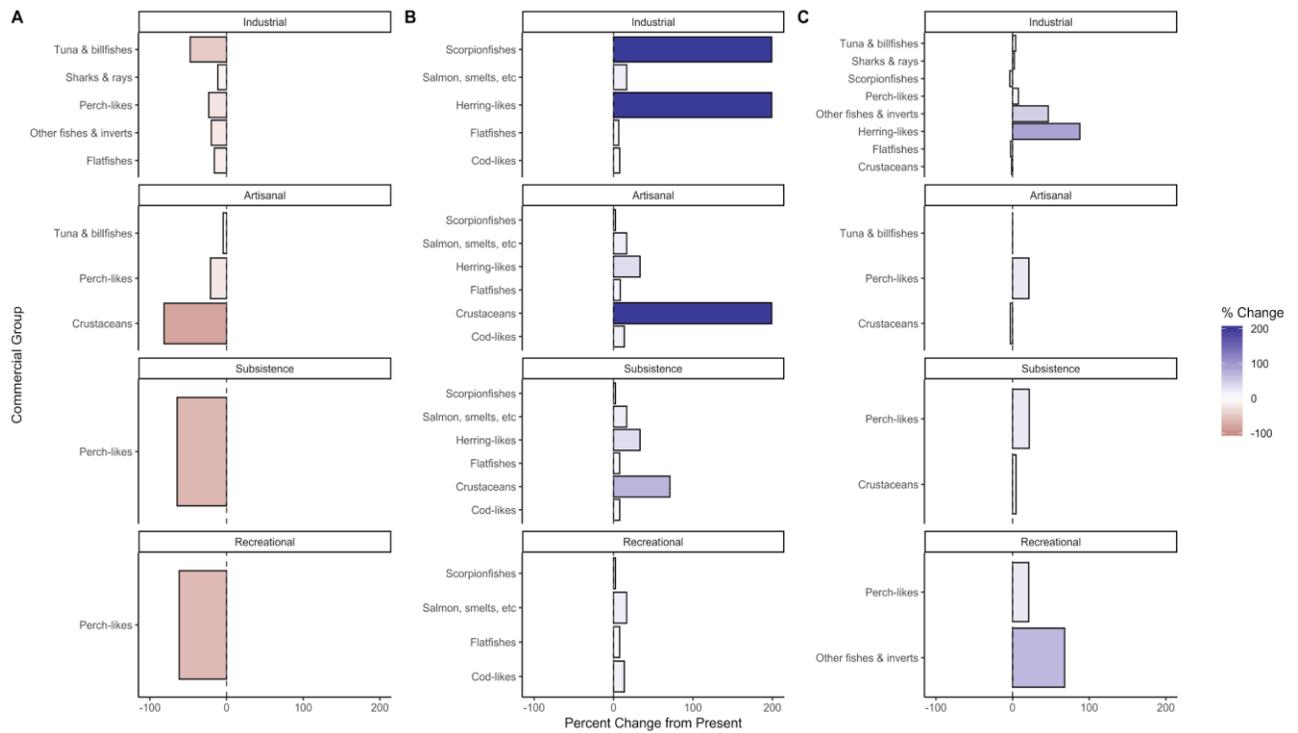

**A) Palau; B) Bering Sea; C) Great Barrier Reef**

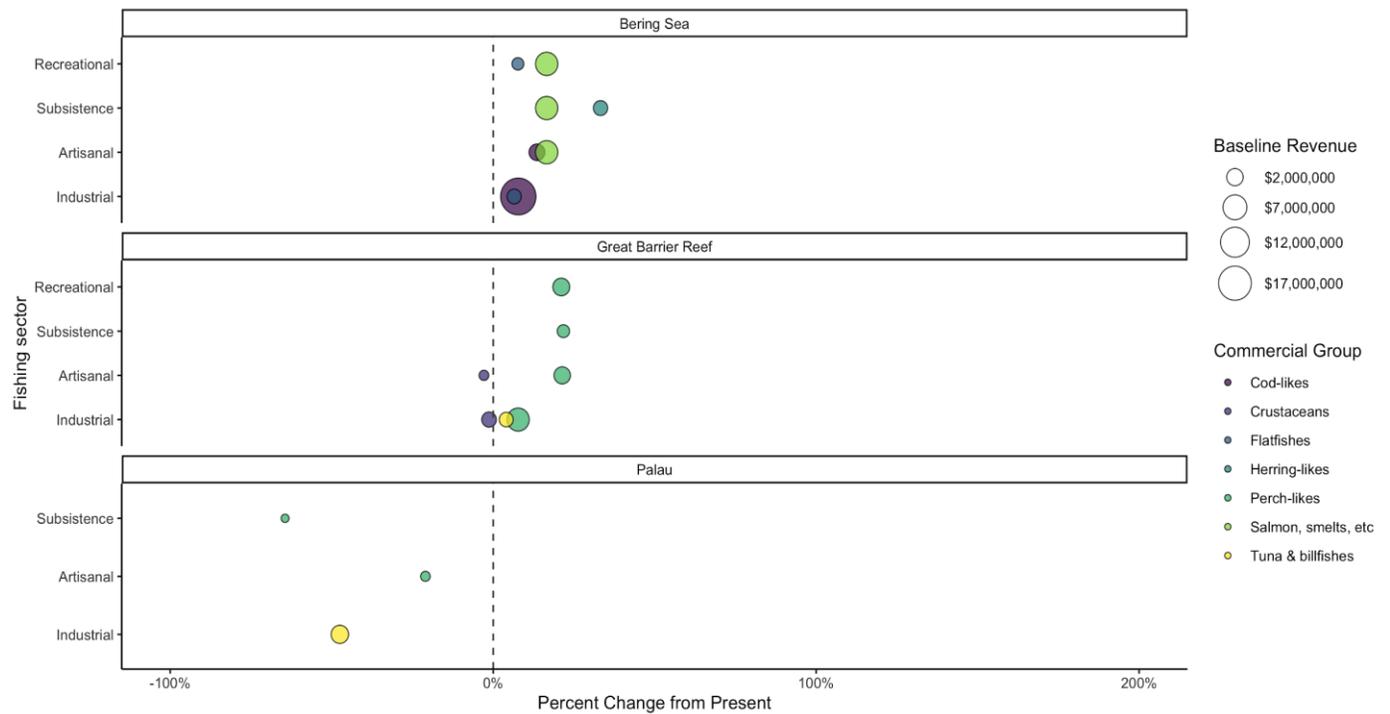

Figure 4. Percent change in revenue from present (1995-2014) to Mid-Century (2040-2060) for most important functional groups for each fishing sector. Bubble size indicates present revenue (real 2010 USD).

### 3.3. Results and Discussion

The findings from the case studies as they were analyzed through the economic methods.

### 3.3.1 Palau

**Biomass:** Palau's MPA will see decreases in biomass for all relevant species and areas in the mid-century. Biomass (within the MPA, adjacent cells, and open ocean) is expected to decrease more towards the equator compared to the northern portion of the MPA. In terms of the percentage change in biomass from present to midcentury, all commercial species are predicted to decrease in biomass with tuna and billfishes decreasing by 48.8% (industrial sector) and perch-likes decreasing by 86.8% for the subsistence sector and 35.5% for the artisanal sector.

**Revenue:** Palau's industrial sector is primarily reliant on tuna-based fisheries, which will see a decrease in revenue by 47.5% towards the midcentury. Palau's artisanal sector is heavily reliant on different reef-associated fisheries, accounting for nearly 90% of their catches and value of their catches. Artisanal revenue from reef-species is expected to decrease by 20.9% by the midcentury. Finally, the subsistence sector is also heavily reliant on reef fishes for ~88% of their catches with some additional catches (5% of their total) from medium sized demersal fishes and by the midcentury revenue from reef fish is expected to decrease by 64.4% for this sector. Thus, these projected declines in reef fishes threaten not only the artisanal sector, but the food security provided by the subsistence sector as well. Due to the greater diversity of species targeted by the artisanal sector, the change in revenues for this sector is less than that experienced by the subsistence fisheries (-20% compared to -65%) where the relative biomass of the perch-like fishes (reef-associated species) is expected to decline by 86.8% (see above). Economic data on the recreational sector of Palau was not available.

### 3.3.2 Great Barrier Reef

**Biomass:** Overall changes in biomass of important marine species for the fishery sector in the Great Barrier Reef will remain somewhat stable until the midcentury. However, the Northern and North-Western parts of the MPA will likely suffer from a decrease of fish biomass while increases are expected in the southern region (Figure 2). In terms of the percentage change in biomass from present to midcentury, all commercial species are predicted to remain close to the baseline (Figure 3.1). This is specifically the case for tuna and billfishes which are projected to experience a 0.7 % increase in biomass within the region by midcentury. The biomass of perch-like species, which is the most important commercial group for the industrial, artisanal, recreational, and subsistence fishery sectors in the region, is also predicted to increase by ~-.01%, ~10%, ~8%, and ~18% respectively. The differences between these sectors can be explained by the types of perch-likes that are most commonly caught by each sector. In fact, the industrial sector fishes mostly for large pelagics, the artisanal sector for large demersals, the recreational sector for the large demersals and medium reef fish, while subsistence fishermen look mostly for medium reef fish and medium pelagics   Finally, crustaceans, which are relevant species for the industrial and subsistence sectors will decrease by about ~ 3% and ~5% respectively (Figure 3.1).

**Revenue:**  Predicted revenue from perch-likes will see an increase under climate change while revenue from crustaceans will decrease for all fishery sectors by the mid-21st century.  The increase of perch-likes is substantial for the entire fishery sector in the Great Barrier Reef area, as this species is the dominant commercial group in this region according to the baseline revenue. The industrial, artisanal, recreational, and subsistence sector will see an increase in revenue from perch-likes of  ~8%, ~22%. ~21%, and ~21% respectively. For the industrial sector, the slight increase in biomass will bring in more revenue from tuna and billfishes (~4%), whereas the revenue from  .Crustaceans, a significantly less important category for the industrial and artisanal

### 3.3.4 Bering Sea

**Biomass:** The overall biomass of fish in the Bering Sea MPA will increase in the mid century (Figure 2). Greater increases occur near the Bering strait, the southern part of the MPA near Nunivak Island, and directly between the Bering strait and Nunivak Island. In terms of the percent change in biomass from present to midcentury, all commercial groups will increase in baseline biomass. The largest baseline biomass in commercial groups is the codlikes, who also have an artisanal sector. The artisanal sector is predicted to increase by a large percentage (15%), which is higher than the industrial sector (6%), most likely because of the difference in starting baseline abundance. The flatfish commercial group will also increase, with the industrial and recreational sector similarly increasing in percent change from present by about 10%. The herring-like commercial group, which is a staple of the subsistence sector, is expected to increase the most by about 30%. Salmon, smelts, etc. commercial group for the recreational sector has a baseline abundance of 500,000 or less and is expected to increase by just under 20% by the mid century.

**Revenue**: Revenue in the Bering Sea will increase in all functional groups and sectors with herring-likes in the subsistence sector increasing by 33%, cod-likes and flatfishes in the industrial sector both increasing by ~ 7%, cod-likes in the artisanal, recreational, and subsistence sector increasing by 16%, and flatfishes in the recreational sector increasing by 7%.

Overall, in artisanal and subsistence fisheries, revenue for crustaceans will increase more than revenues for any other groups, followed by herring-likes. The industrial fisheries will have the greatest increases in revenue, both in total amount and percentage, though the impact is most profound in scorpionfishes and herring-likes. The increase in revenue for salmon species is moderate and consistent across sectors, and the baseline revenues across sectors are also very similar. Revenue in the recreational sector in the Bering Sea will see the most modest increase in all functional groups. The differences in revenue changes across sectors for a single functional group can be explained by two reasons. First, industrial fishing covers a much larger area than artisanal, recreational, or subsistence fishing, so the increase in biomass may differ (as mentioned in the biomass paragraph above). Second, the specific key species for different sectors may be different, even though they are in the same group. Revenues for these species may be quite different from each other, leading to the differences in revenues across sectors for each functional group.

**3.4 Cross-study comparison**

Our model predicts that out of the three MPAs we studied, the Bering Sea will see the most gains both in terms of biomass and revenue. This is mainly because the Bering Sea is located in a higher latitude, and thus has a cooler climate than the GBR and Palau, and changing ocean conditions will cause species to shift distributions into cooler climates. Conversely, the Palau National Marine Sanctuary resides close to the equator, where ocean warming will most likely drive species away from these areas towards the poles. The GBR represents an intermediate case, where certain species will shift into the area from the tropics, while others will shift away from the area towards the poles. Thus, the predicted decreases in revenue and biomass are greatest in Palau, and less severe in the GBR, while the Bering Sea is expected to experience increases in biomass and revenue. In all cases, the magnitude of change in biomass and revenue will differ across sectors and functional groups.

# 5. Conclusion

We found that fish biomass near Palau, our case study near the equator, is predicted to sharply decline, while fish biomass near the Great Barrier reef, our tropical case study further from the equator, is predicted to slightly decline, and fish biomass in the Bering Sea Region, our third case study, is predicted to increase. Our findings show that as expected, biomass and revenue will increase in colder latitudes, although sectors will be impacted differently. This type of analysis can be used in other locations to also study how the communities around marine protected areas will be affected by warming ocean waters. What has emerged is the understanding how different sectors will be impacted by the species. This paper directs towards effective policy choices to protect ecosystems and promote user outcomes. Recommendations for future research include local drivers such as fisheries, tourism, indigenous populations on the likelihood of successful MPA governance. Research should also be conducted on the socio-economic benefits of small- and medium-size MPAs under climate change, and networks. Based on our findings, future research should also investigate the fish groups in this paper to see if MPAs under climate change are more effective for some species than for others. Finally, regional climate models that incorporate local ecological phenomena are needed to further understand the impacts of climate change on these specific areas.

These sectors demonstrate the important role of socio-economic context to managing MPAs and their surrounding waters under climate change. Through understanding where the fisheries currently fish and are able to fish in the future, and which species or fish groups will increase in biomass, regulators can prepare for climate change through regulations that decrease pressure on species that will decrease in biomass under climate change. This method can be used by managers in other areas to predict economic changes under climate change. The success of MPAs is connected to the socio-economic surroundings of the MPA. Fisheries, tourism, and presence of indigenous communities are critical drivers that can help or hinder the success of MPAs for conservation.

*Acknowledgements:*
*Thank you to Tim Cashion, Juliano Palacios Abrantes, Sarah Roberts, Anne Mook and Tu Nguyen for their valuable work on this project. This work was supported by the National Socio-Environmental Synthesis Center (SESYNC) under funding received from the National Science Foundation DBI-1639145.*